# A Novel UV Photon Detector with Resistive Electrodes

J.M. Bidault[1], P. Fonte[2,3], T. Francke[4], P. Galy[1], V. Peskov[1,*], I. Rodionov[5]

1 Pole Universitaire Leonardo de Vinci, Paris, France
2 Instituto Superior de Engenharia de Coimbra, Coimbra, Portugal
2 LIP-Laboratório de Instrumentação e Física Experimental de Partículas, Coimbra, Portugal
4 XCounter AB, Danderyd, Sweden
5 Reagent Research Center, Moscow, Russia

## Abstract

In this study we present first results from a new detector of UV photons: a thick gaseous electron multiplier (GEM) with resistive electrodes, combined with CsI or CsTe/CsI photocathodes.

The hole-type structure considerably suppresses the photon and ion feedback, whereas the resistive electrodes protect the detector and the readout electronics from damage by any eventual discharges.

This device reaches higher gains than a previously developed photosensitive RPC and could be used not only for the imaging of UV sources, flames or Cherenkov light, for example, but also for the detection of X-rays and charged particles.

[*] Corresponding author: V. Peskov, CERN, Geneva 23, Switzerland (e-mail: vladimir.peskov@cern.ch).

## I. Introduction

There has been recently considerable interest on Ultra Violet (UV) imaging detectors, with have applications on homeland security [1]and on the detection of small forest fires or corona discharges at large distances [2].

The most commonly used UV imaging detectors are image intensifiers combined with narrow-band filters transparent in the 240 to 280 nm band [3]. At this wavelength range, called "solar blind", the daytime detection of UV light occurs with practically no background of solar radiation, which is absorbed by the ozone layer in the upper atmosphere and does not reach the Earth surface. Thus the UV light emitted from fire or electrical discharges may be detected with a very high signal to background ratio. For example, a small forest fire can be detected by such a device up to a distance of several kilometers [4].

Recently we have developed a new type of detector of single UV photons: Resistive Plate Chambers (RPCs) combined with CsI or CsTe photocathodes [5]. These RPCs operate at high gas gains (up to $10^6$) and the resistive electrodes to protect the readout electronics from damage by possible discharges. With a position resolution of 30 μm for single photoelectrons these could also be used in UV imaging applications, like RICH or in the visualization of flames air. Preliminary tests on flame detection show that our RPCs have the same sensitivity as the solar-blind MCP-based image intensifiers [6]. This is because the image intensifiers are combined with narrow band filters having a transmission close to 10% at the peak of the transmission curve, while our RPCs operate in a photon counting mode without any filters. Even better sensitivity may be expected if the RPC could be combined is with CsTe photocathodes, whose quantum efficiency overlaps better with the flame emission spectra [5]. First results show that RPCs combined with CsTe photocathode can operate stably, but the maximum achievable gain was limited to approximately $10^3$ owing to feedback problems.

In this work we developed an innovative hybrid device: a thick GEM-like [7] electron multiplier (TGEM [8],[9],[10]) with resistive electrodes, which we call “Resistive Thick GEM - RETGEM”. The hole-type structure considerably suppresses the photon and ion feedback, while the resistive electrodes protect the detector and the readout electronics from damage by any eventual discharges.

## II. Detectors and Experimental Set Up

A schematic drawing of the detector is shown in Fig.1.

It contains a semitransparent CsI photocathode, 20 nm thick, deposited on a $CaF_2$ disc and placed 7 mm away from the electron multiplier. This is constituted by a hole-type structure composed of a G-10 plate either 2 mm thick with holes of 1 mm diameter and 1.6 mm pitch (see [8]) or 0.4 mm thick with holes of 0.3 mm diameter and a pitch of 0.7 mm. The Cu coating of the G-10 plate was almost fully removed (except in some very small random spots) and both surfaces were covered with a graphite paint typically used in RPCs [11]. The graphite coating around the holes on a distance of approximately 0.25 mm was removed by a drill of a larger diameter.

The readout plate was segmented in metallic pixels with 1.27 mm pitch (similar to the one described in [9]) and placed 0.4 to 1 mm below the RETGEM anode. To simplify the readout electronics, in most of measurements the pixels were electrically connected in rows and amplifiers were connected to each row, providing a 1D projection of the image.

Two RETGEM operation modes were considered: with the gas gain occurring inside the holes – “mode A” (Fig. 2a) - and with the gas gain occurring in the gap between the RETGEM anode and the readout plate – “mode B” (Fig. 2b).

The ensemble was installed inside a gas-tight chamber equipped with a $CaF_2$ window, as shown in Fig. 2. As UV sources we used candles and a pulsed (few ns) UV lamp. In some tests a quartz lens was used to project the image of the UV sources on the photocathode. For gain measurements, down to unit gain, the $H_2$ pulsed lamp was attached to the chamber window via an Ar-flushed tube.

For rate capability measurements we used an X-ray generator and replaced the photocathode by a wire mesh that defined an X-ray conversion gas volume above the RETGEM.

Comparative measurements were performed with a TGEM, a photosensitive RPC and an UV image intensifier developed by Reagent (described in [6]). To make comparisons easier, these tests were done in the same gas mixtures as we used before for the photosensitive RPCs and TGEMs: Ar+5%$CH_4$, Ar+5% isobutane [12] and He+0.8%$CH_4$+EF* [5] at a total pressure of 1 atm.

## III. Results

In Fig. 3 we show the effective gain as a function of the voltage applied across the TGEM and RETGEM detectors operating in mode A. The effective gain is the ratio of the primary charge to the charge collected on the readout electrode; therefore it is somewhat lower than the total multiplication. It is apparent that, owing to the resistive coating, the RETGEM allows much larger gains to be achieved.

In the same figure one may appreciate that at the larger voltages RETGEM develops a mixture of pulses with very different amplitudes, represented by the open and filled symbols. This corresponds to the appearance of mild discharges (commonly denoted by “streamers” in RPC terminology) whose maximum current is limited by the resistive electrode. If not too frequent, these do not hinder the normal operation of the detector and not propagate to the readout plate even if a few kV were applied between the anode of the RETGEM and the readout plate. This is in contrast to the behavior of GEM or TGEM detectors, where discharges in holes may easily propagate to the readout plate (see [13] for more details).

As the voltage is further increased, permanent discharges may appear in some holes. These are not destructive but will reduce the effective applied voltage and normal operation becomes impossible.

The performance of the RETGEM combined with the CsI photocathode is also shown in Fig. 3.

Fig. 4 shows the effective gain as a function of the voltage applied between the lower electrode of the RETGEM and the readout plate (mode B – see Fig. 1b) and, for comparison, for the RPC described in [5], both combined with a semitransparent CsI photocathode. Again, it is clear that RETGEM reaches higher gains than the photosensitive RPC owing to the photon feedback suppression provided by the hole-type structure.

Fig. 5. shows gain vs. counting rate curves for TGEM and RETGEM detectors illuminated with X-rays. The gain of the TGEM practically does not change with rate, whereas the RETGEM gain drops with rate in a way qualitatively similar to low resistivity RPCs [14]. At counting rates larger than $10^6$ Hz/$cm^2$ some kind of glow discharges may appear in the holes

---

* Ethyl Ferrocene.

or in the gap between the RETGEM and the readout plate. However, these were mild and harmless to the readout electronics.

One of the goals of this work was to investigate the possible application of RETGEM to flame imaging. Fig. 6 shows the 1D projection of the candle flame image obtained with the set up shown in Fig. 2. For this particular measurement the candle was placed 1.5 m away from the detector and the counts were accumulated during 0.1 s. When the candle was placed 30 m away it produced counting rates on the order of a kHz, to be compared with a background of less than 10 Hz in a room fully illuminated by natural or fluorescent lights

We also performed quantitative comparisons between the RETGEM and a single-wire counter with CsI photocathode previously developed by us [15]. Both detectors had almost the same sensitivity to the flame and could operate in fully illuminated rooms. However, in sunlight conditions the signal to noise ratio $k=S/N$ (where $S$ is the excess counting rate produce by the flame and $N$ is the counting rate without the flame) was small ($k \ll 1$). However, for RETGEM the signal to noise ratio was much higher: $S/N \sim kn$, where $n$ is the number of pixels in the readout plate. This was because the image of small flames occurring far from the detector occupied one or a few pixels, whereas the diffuse sunlight was distributed over all pixels. Moreover, direct sunlight was also projected on a few pixels and would not interfere with the image of the flame. If necessary, when operating in mode B, the pixels irradiated by direct sunlight could be electrically disconnected (left floating), drastically reducing the local gas gain and sparing the CsI photocathode.

First tests of RETGEM combined with CsTe/CsI photocathode [6] were also performed, yielding the results shown in Fig. 4. A maximum gain larger than $10^4$ was reached, which is almost an order of magnitude larger than what is possible with an RPC combined with CsTe photocathode [5].

Finally we performed preliminary comparisons of the RETGEM characteristics with an MCP-based image intensifier [6]. Both detectors had the same sensitivity for flames occurring in rooms, but the image intensifier had better signal to noise ratio for flames detected in sunlight conditions. One of reason for this was that the image intensifier had a much larger number of pixels. Probably RETGEM combined with micropixel redout, such as MediPix, will be able to compete with image intensifiers for outdoor UV detection.

## IV. Conclusions

The development described in this work, the Resistive Thick GEM (RETGEM), suggests that RETGEM is a new promising gaseous detector which combines the spark protective properties of the RPC with the unique properties of the hole-type structures/GEM, such as the possibility to extract electrons from the holes, directing them to another multiplication structure or to the readout electrodes and the efficient suppression of photon and ion feedback. This opens the possibility to combine RETGEM with various photocathodes and use it as an UV imaging device. Probably, after further developments, RETGEM-based imaging devices may become compact and simpler alternatives to the UV image intensifiers used today.

When combined with a CsI photocathode, RETGEM could be particularly valuable in RICH detectors, as it allows large gas gains over large areas and it is protected from the discharges that might be triggered by highly ionizing particles. Obviously, RETGEM could be also used for the detection of charged particles.

As in the case of RPCs [16],[17], by optimizing the resistivity of the coating one can probably the rate characteristics of the RETGEM may be considerably improved, further enlarging the application fields.

Further developments may include the manufacture a microhole resistive electrode structure or resistive micromesh (as was suggested in [18]) and this in turn will open new applications. For example, such a resistive micromesh could be an interesting alternative to MICROMEGAS or GEMs combined with MediPix readout, as the resistive mesh will protect the Medipix from destructions in the case of occasional discharges.

**References:**

[1] http://www.sbuv.com/index.html
[2] http://www.daycor.com/
[3] G.F. Karabadzhak et al., "Mir-based Measurements of the UV Emission from Rocket Exhausted Plumes Interacting with the Atmosphere" Report at the 38 Aerospace Science Meeting and Exhibition, Reno, Nevada 2000, Preprint AIAA 2000-0105.
[4] http://www.daycor.com/fire.html
[5] P. Fonte et al., Nucl. Instr. and Meth. , A533, 2005, 30.
[6] I. Rodionov et al., Preprint Physics/0511212, Nov. 2005.
[7] F. Sauli, Nucl. Instr. and Meth. A 386, 1997, 531.
[8] L. Periale et al., Nucl. Instr. and Meth. , A 478, 2002, 377.
[9] J. Ostling et al., Nucl. Instr. and Meth. , A 525, 2004, 308.
[10] R. Chechik et al., Nucl. Instr. and Meth. A 535, 2004, 303.
[11] Supplied by C. Gustavino , Gran Sasso Lab.
[12] J. Ostling et al., IEEE Trans. Nucl. Sci., 50, 2003, 809.
[13] C. Iacobaeus et al., IEEE Trans. Nucl. Sci., 48, 2001, 1496.
[14] P. Fonte et al., Nucl. Instr. and Meth. A431,1999,154.
[15] P. Carlson et al, NIM 505 (2003) 207.
[16] P. Fonte et al., Nucl. Instr. and Meth. A 431, 1999, 154.
[17] T. Francke et al., Nucl. Instr. and Meth. A 508, 2003, 83.
[18] Van Der Graaf et al., "Tracking cosmics: recent results from MICROMEGAS-covered MediPix2 pixel CMOS readout circuit in a mini-TPC" Report at the 7th International Conference on Position-Sensitive Detectors, Liverpool, September 2005.

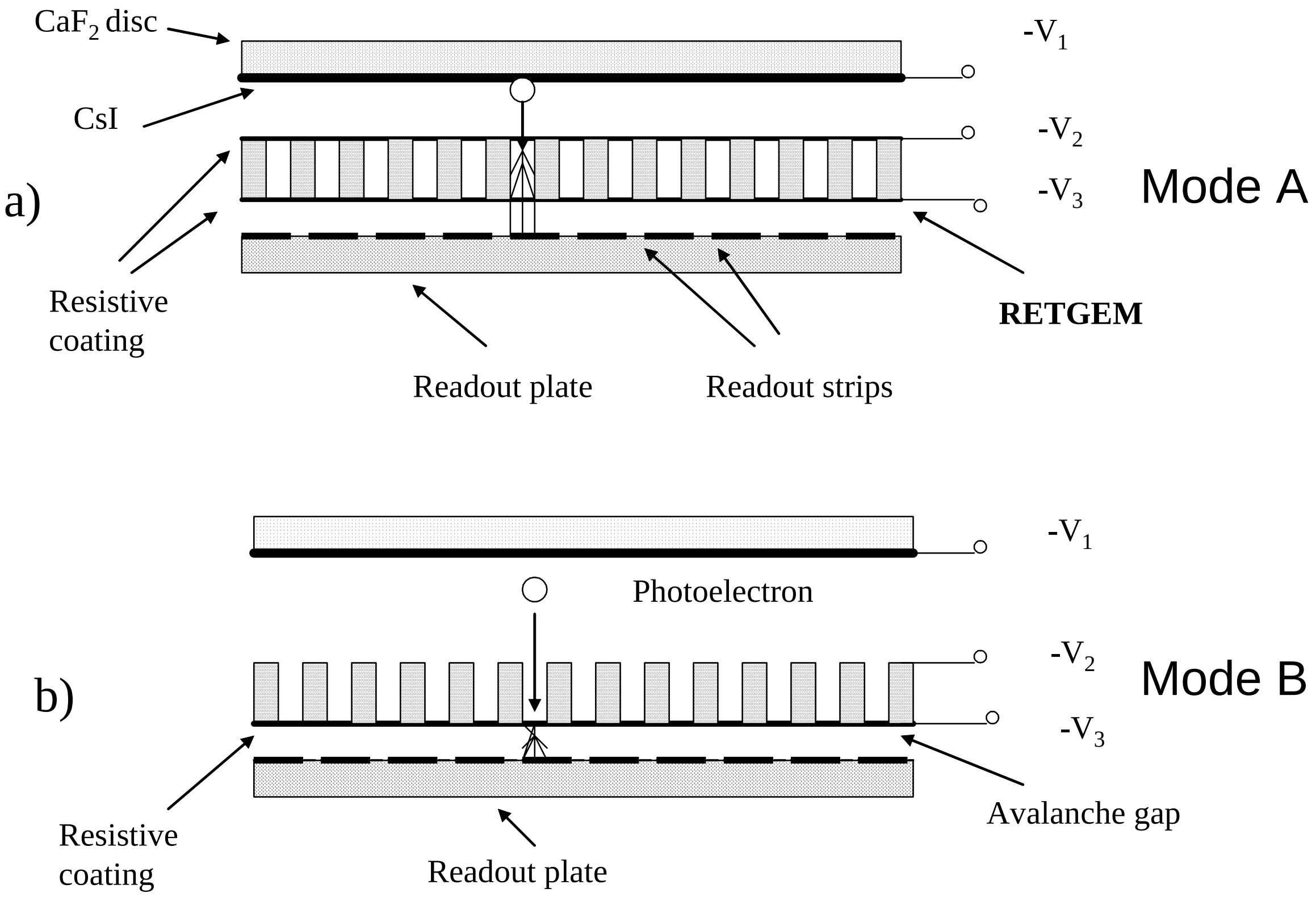

Fig.1. A schematic drawing of the Resistive Thick GEM (RETGEM) detector studied in this work, illustrating both operating modes.

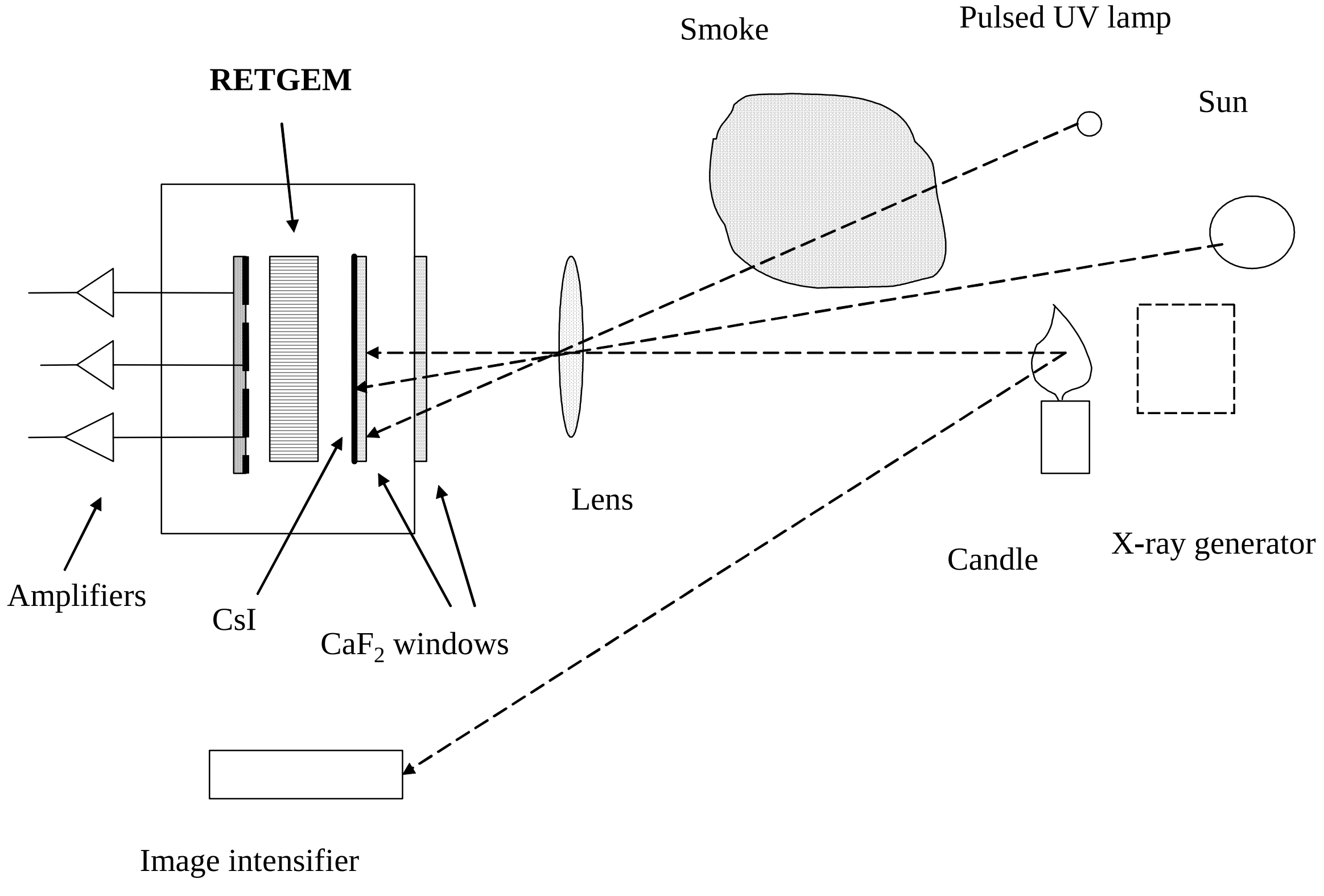

Fig. 2. A schematic drawing of the experimental set up used for the RETGEM tests.

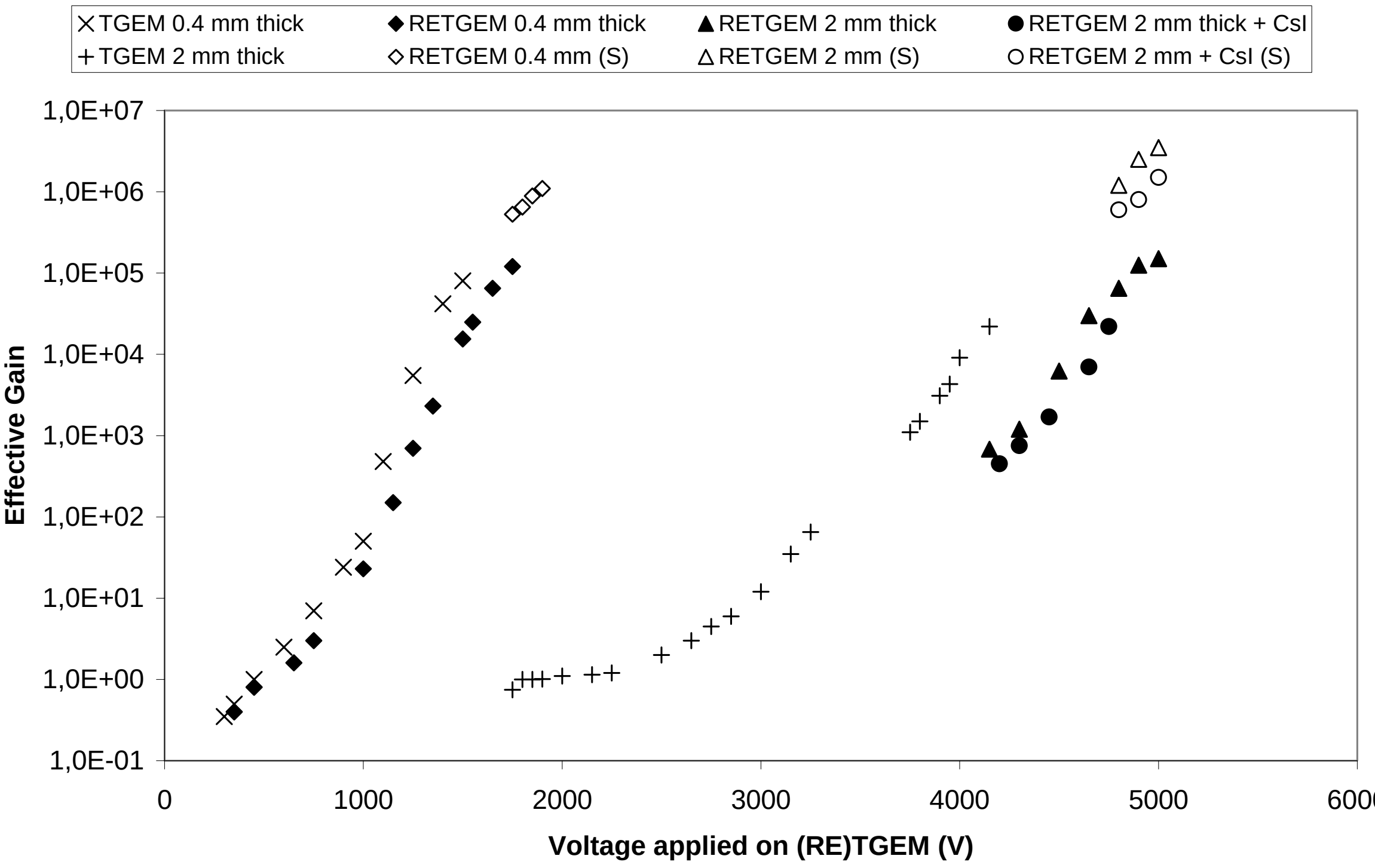

Fig. 3. Gain vs. voltage curves for Thick GEM (TGEM), Resistive Thick GEM (RETGEM) and RETGEM combined with a CsI photocathode. Tests of 0.4 mm thick detectors were done in Ar + 5%$CH_4$ and 2 mm detectors were tested in Ar + 5% isobutane. The mark “(S)” in the legend indicates that these points correspond to “streamer” discharges.

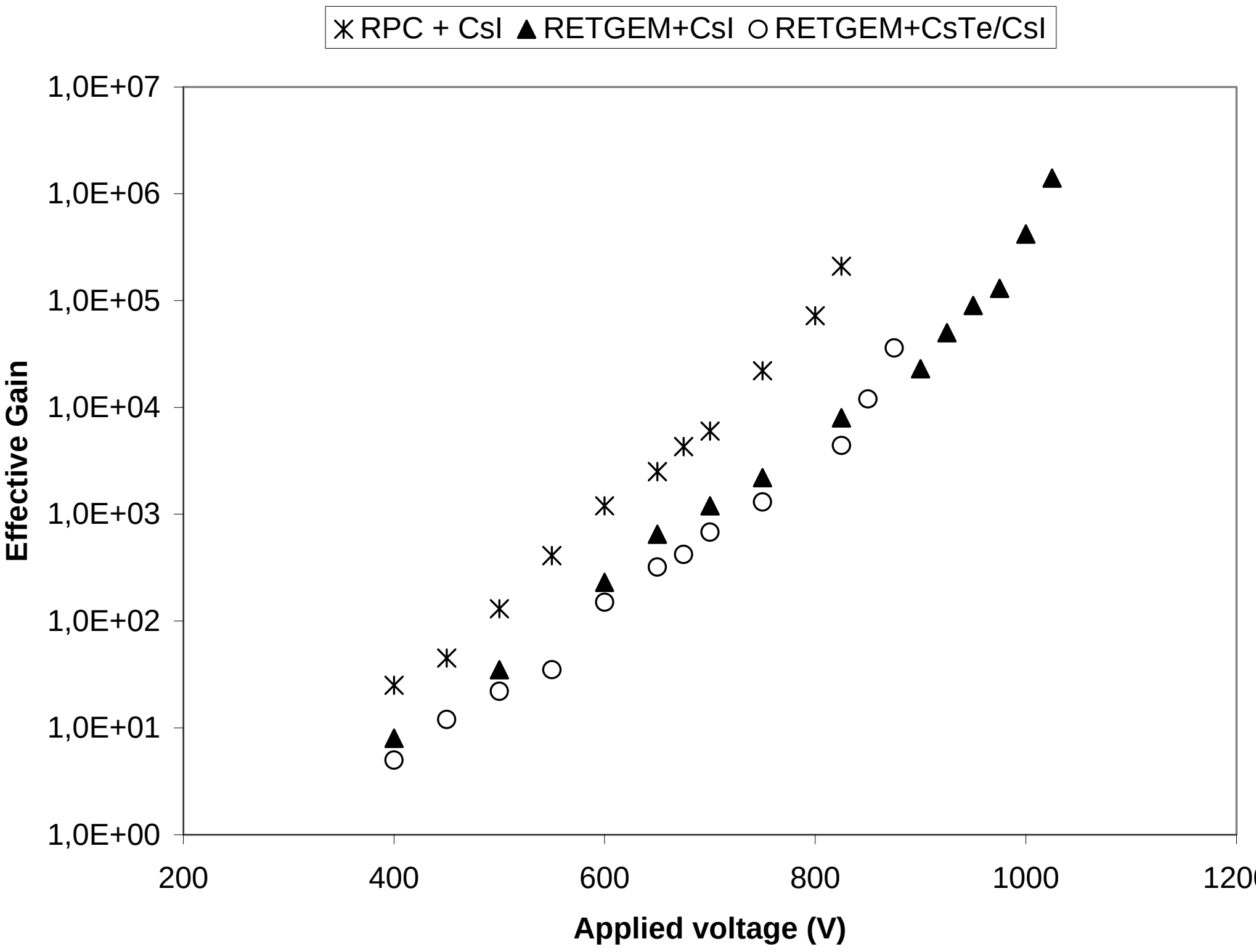

Fig. 4. Gain vs. voltage curves for RPC and RETGEM operating in mode B, both combined with semitransparent CsI or CsTe/Csi [6] photocathodes. All measurements were done in a He+1%$CH_4$+EF gas mixture.

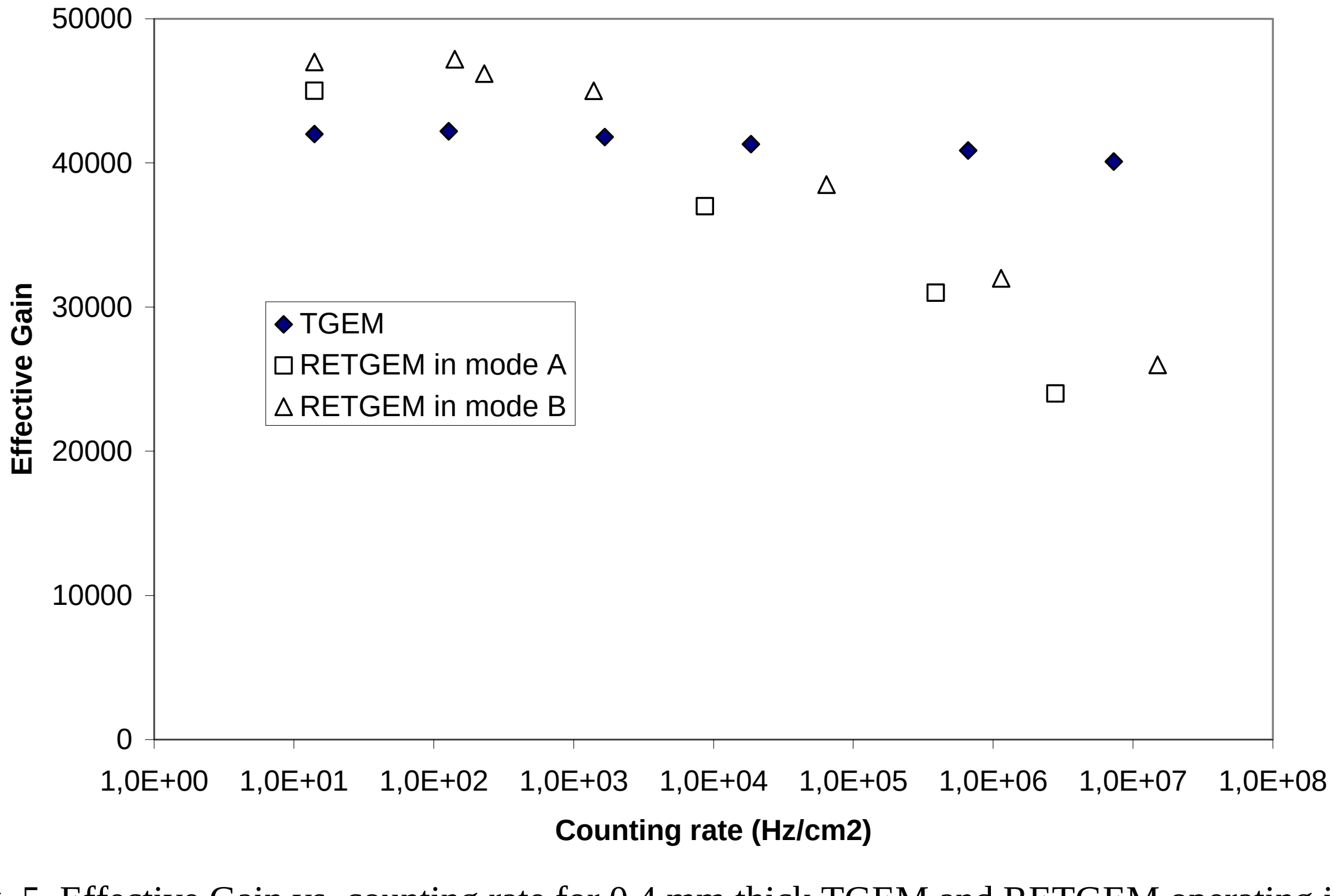

Fig. 5. Effective Gain vs. counting rate for 0.4 mm thick TGEM and RETGEM operating in modes A and B.

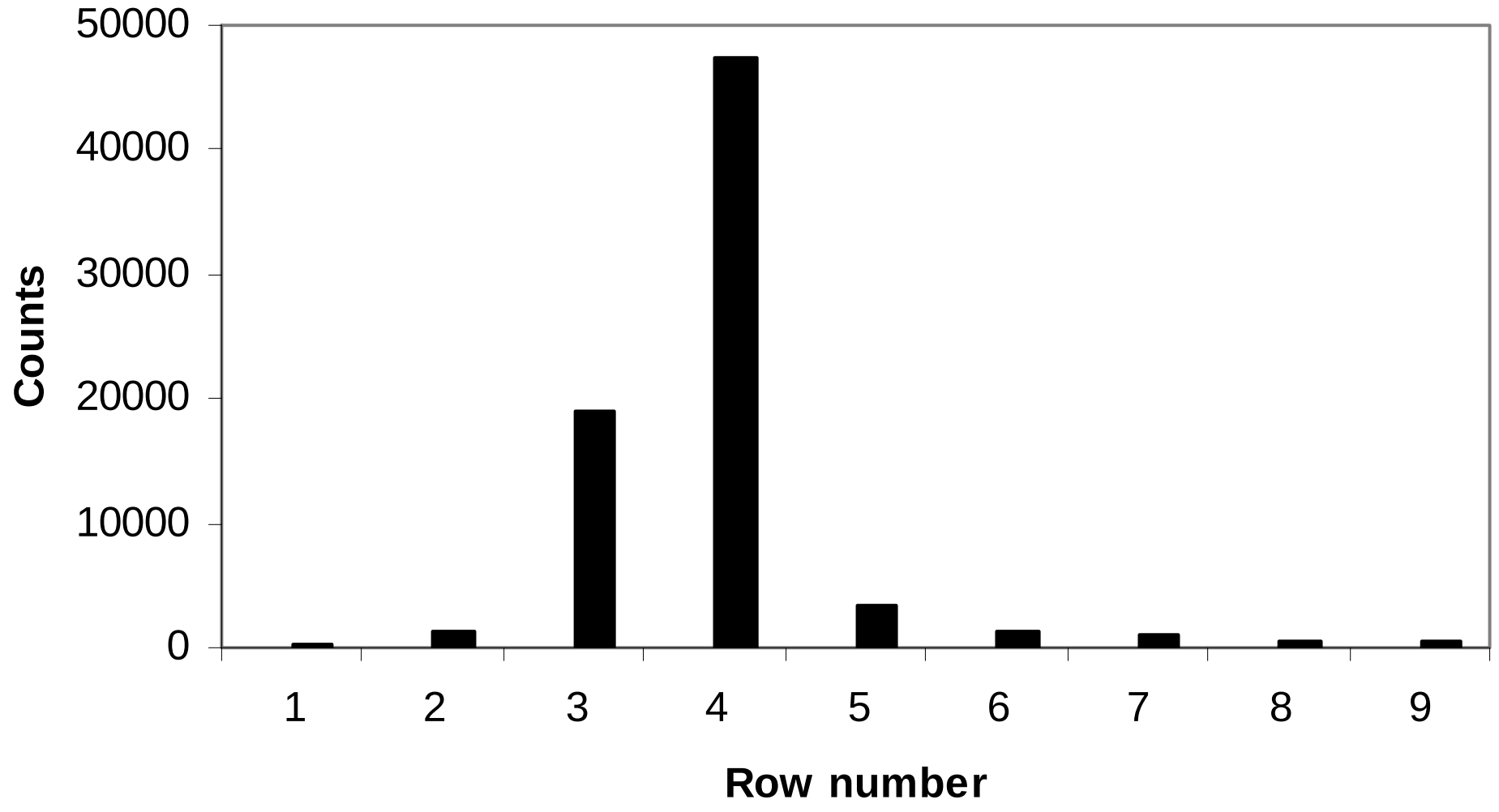

Fig. 6. A digital image of a candle flame obtained with the RETGEM combined with a semitransparent CsI photocathode. The rows are separated by 1.27 mm.